# Reaction $^{nat}$Yb+$^{48}$Ca$\rightarrow$$^{217}$Th+3n: auto calibration process for DSSSD detector application (particular case)

Yu.S.Tsyganov

*FLNR, JINR, 141980 Dubna, Russia*

[tyura@sungns.jinr.ru](mailto:tyura@sungns.jinr.ru)

**Abstract**

The Dubna Gas-Filled Recoil Separator (**DGFRS**) is the mostly effective setup in use for the field of synthesis of superheavy elements. Application of DSSSD detector provides more precise energy and position detection for both implanted recoils and their alpha decays. To minimize contribution of background products the method of "active correlation " is applied at the DGFRS. To apply this method a precise calibration for 48 front strips is strongly required. Method of three peaks auto calibration is proposed with a first approximation step with 9.26 MeV $^{217}$Th peaks. Examples of α-decay original spectra and spectra after filtration procedures are presented.

## 1. Introduction

A series of successful experiments performed at the gas-filled separator of recoil nuclei (DGFRS) over the past few years was made possible by a high degree of automation of search for rare decays of superheavy nuclei (SHN) [1-4]. Namely with the DGFRS more than fifty isotopes of SHN were discovered. To detect rare alpha decays in a background free mode "active correlations" method was developed and successfully applied [5-8]. These experiments provided clear manifestation of so called "island of stability" discovery.

From the viewpoint of applied mathematics methods and algorithm application one can mention for items for such on-line experiments, namely:

- Estimation of the statistical significance value (probability, expectation) for each detected multi chain event;
- Algorithm of quick sort (real-time) of correlated events like ER-α (recoil-alpha) for implanted into silicon detector nuclei;
- Auto calibration of multi strip detector;
- Nearest future trends of applications under condition of ultra intense heavy-ion beams.

Because of items 1 and 2 are described exhaustively in the Ref's [5-8, 9-11], the last two items are in focus of the present paper.

Of course, transport modeling of the setup is also applied mathematics question, but it far apart from the on-line experiments area.

## 2. Express method for 48x128 strip Micron Semiconductors DSSSD detector auto calibration.

The specific of alpha-decay spectrum from the complete fusion heavy ion induced nuclear reaction

$^{nat}Yb + ^{48}Ca \rightarrow ^{217}Th + 3n$ is a relatively short life time for that isotope, namely 237 μs. It means that it is easy to identify that isotope via searching for ER-α correlation time-energy-position. The presence of a spectral line at 9262 KeV is among the mentioned specific. This line is easily to identify automatically and may be used for calibration to a first approximation with a single line ($a_0$=9261/center of peak gravity). In the second approximation two others right hand lines 8700 and 7923 are taken into account and Least Square Method (LSQ) is applied for these three lines.

```
// ---- Builder C++ code fragment

const float e0=7923.0; const float e1=8700.0;

For (i=0; I < 3 ; i++) {

if  (enrg_1[i] >e0-dddw && enrg_1[i] < e0+dddw) enrg_1[i]=e0;

 if (enrg_1[i] >e1-dddw && enrg_1[i] < e1+dddw) enrg_1[i]=e1;

}
```

Below, the result of comparison for this auto calibration method with respect to the precise off-line eight peaks LSQ method (lines:6040.,6143.8, 6731 , 6899.2,7137.,7922.,8699.,9261 ) is presented as deviations for the energy vicinity of 9 MeV value both in text block and in the Fig.1.

{1 ,1;  2, 5; 3, -7; 4, -3; 5 -2;  6, 1; 7, 1; 8, 1; 9, 0 ; 10, -1; 11, -3 ; 12 ,-3 ; 13, -4; 14, -3; 15, -1; 16, 0 ; 17, 1 ; 18, -3 ; 19, 0; 20 ,-2; 21, 0; 22, 3 ; 23, -2; 24, 0 ; 25, 0 ; 26, 0 ; 27, -4; 28, 0 ; 29, 0 ; 30, 0; 31, 0; 32, 0; 33, 2; 34, -2; 35, -2; 36, 0; 37, 0; 38, -4; 39, 0 ; 40, -1; 41, -1; 42, 2; 43, -1; 45, -1 ;46 ,3 ; 47 ,0  }.

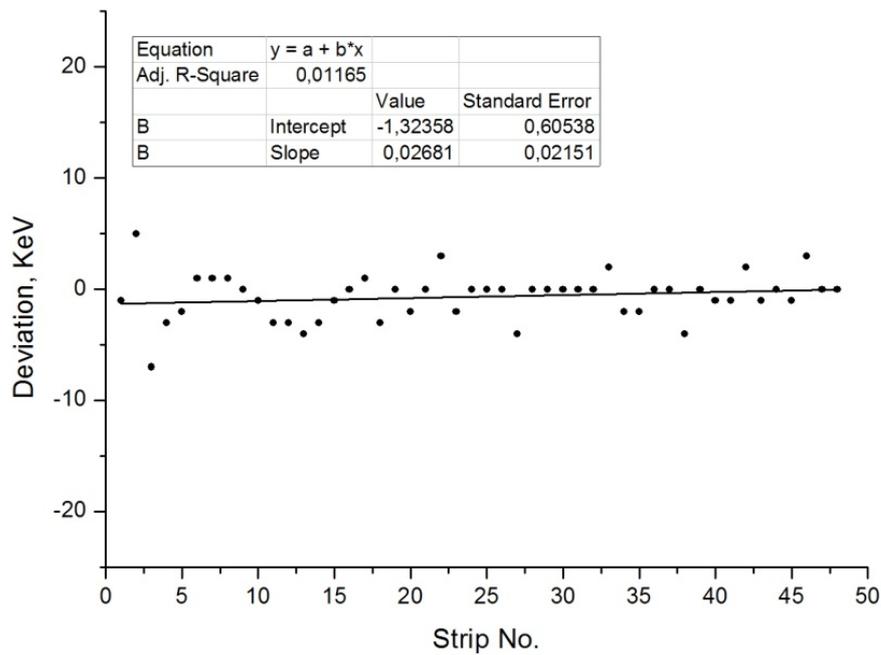

**Fig.1** Deviations from the precise calibration in the ~9 MeV energy signal vicinity.

First digit in the presented pairs denotes strip number, the second one – deviation in KeV's.

When one use this calibration procedure for masses around 300 a.m.u. a small additional systematic error can be arisen [12] as:

$$\mu = \frac{e^\alpha_{SHE}}{e^\alpha_0} \approx \frac{1+\eta \cdot \frac{m}{M_0}}{1+\eta \cdot \frac{m}{M_{SHE}}}.$$

Here, indexes "0" and "SHE" are corresponded initial (according to calibration procedure) and corrected energies, respectively ($M_0$=217, $M_{SHE}$ ≈300).

### 3. High intensity beam nearest future application

With commissioning high intense DC-280 heavy ion cyclotron [13] in a nearest future (~2017-2018; ~10 pμA for $^{48}$Ca beam!) problem with non-definite start for ER-alpha sequences will be quite reasonable. It was V.B. Zlokazov who first recognized the importance of the theoretical approach to the non-definite mother–daughter nuclei relationship. He epitomized mathematically an equation system to search for an actual life time value [14]. A more simplified mathematical approach for two candidates for recoil (EVR) was reported in [15] in the form of a transcendental equation relative to the actual life time parameter, namely:

$$\tau = \frac{t_1(1 - e^{-\frac{t_1}{\tau}}) + t_2(1 - e^{-\frac{t_2}{\tau}})}{2 - e^{-\frac{t_1}{\tau}} - e^{-\frac{t_2}{\tau}}}$$

Here, $t_1$ and $t_2$ are the detected life times, and $\tau$ is the actual (virtual) parameter to be found in a real-time mode. According to that reason procedure of $\tau$ finding should be incorporated into the data acquisition code (C++ Builder 6.0, REDSTORM.exe, Windows [16] ). Below, in the Fig.2 Newton method solution of the above mentioned equation is shown. It takes about ~3 μs to provide of units of percents accuracy that is quite sufficient for active correlations technique application. Note, in the case of one considers $\tau_0 = \sqrt{t_1 t_2}$ value for a first approximation only one-two iteration is required for several percents accuracy.

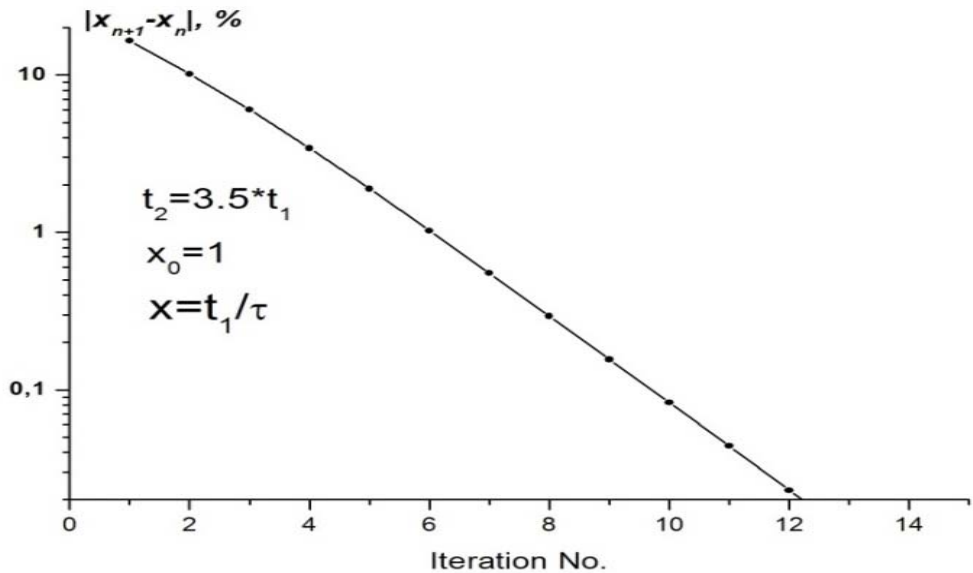

**Fig.2** The dependence of the solution accuracy against number of iterations (i3-2100 CPU@3.10 GHz). One iteration time interval is equal to about 1.4 μs.

In a more common case values registered energy and TOF/ΔE (time of flight/ losses in pentane gas for START and STOP low pressure proportional counters) signal shape may be taken into account under formation weight function for recoil signals "1" and "2". Therefore, the required condition for a beam stop will be in the form of equation:

$$\tau(E_1^{REG}, E_2^{REG}, TOF_1, TOF_2, \Delta E_1, \Delta E_2, t_1, t_2) \leq t_{SET}.$$

Here, $t_{set}$ is a pre-setting by the experimentalist time parameter for ER-α correlation chain, $E_{1,2}$- registered ER's energies, $t_{1,2}$- registered times between alpha decay and recoils "1" and "2", respectively.

## 4. Summary

Express method for auto calibration of DSSSD detector is developed using heavy ion induced complete fusion $^{nat}Yb+^{48}Ca$ reaction. Reasonable scenario to modify active correlation method in the case of ultra high beam intensity is considered.

Only a few microseconds require additionally making the appropriate procedure located within the body of data acquisition code REDSTORM.exe (Windows, C++).